\documentclass[prl,aps,twocolumn,amsmath,amssymb,a4paper,showpacs]{revtex4}	
\usepackage{graphicx}
\begin{document}
\title{An Optical Clock with Ultracold Neutral Atoms}
\author{G.~Wilpers, T.~Binnewies, C.~Degenhardt, U.~Sterr, J.~Helmcke, F.~Riehle}
\affiliation{Physikalisch-Technische Bundesanstalt, Bundesalle 100, 38116 Braunschweig, Germany}

\date{\today}
\begin{abstract}
We demonstrate how to realize an optical clock with neutral atoms that is competitive to the currently best single ion optical clocks in accuracy and superior in stability. 
Using ultracold atoms in a Ca optical frequency standard we show how to reduce the relative uncertainty to below $10^{-15}$.
We observed atom interferences for stabilization of the laser to the clock transition with a visibility of 0.36, which is 70\% of the ultimate limit achievable with atoms at rest.
A novel scheme was applied to detect these atom interferences with the prospect to reach the quantum projection noise limit at an exceptional low instability of $ 4 \times 10^{-17}$ in 1~s.
\end{abstract}
\maketitle

Progress of science and technology has been closely connected to the development of sophisticated methods to measure time and frequency. 
From the highly developed pendulum clocks of around 1900 to today's most advanced Cs atomic clocks, realizing the SI time unit, the increase in relative accuracy from $10^{-7}$ to $10^{-15}$ was mainly made possible by an increase in the operational frequency from the Hz to the GHz range.
With optical comb generators based on femtosecond lasers \cite{hol01c} now optical frequencies can be connected to radio frequencies with an uncertainty below $10^{-18}$ \cite{ste02a}, enabling optical clocks at frequencies that are five orders of magnitude higher than the frequency of the Cs transition.

With this 'clockwork' available the performance of an optical clock primarily depends on the 'pendulum', i.e. on the atomic or molecular transition, and equally important on the method that is used to interrogate the 'pendulum' with the least noise while keeping all disturbances at a minimum. In an optical clock this corresponds to probing the transition at the best possible signal-to-noise ratio (SNR) and stabilizing the laser to the true undisturbed line center.

The achieved uncertainties of optical clocks based on single ions of $\lesssim 10^{-14}$ (Hg$^+$ \cite{ude01}, Yb$^+$ \cite{ste01a}) and on large numbers of neutral absorbers of $\gtrsim 10^{-14}$ (Ca \cite{ude01,wil02}, H \cite{nie00}) are still about an order of magnitude worse than the best Cs atomic clocks \cite{cla96,wey01,jef02}. In contrast to the single ion standards, where the absorber is confined to a small volume, the neutral atom standards are ultimately limited by the residual velocity of the absorbers, but, on the other hand, benefit from the higher signal to noise ratio due to the large number of atoms which can lead to exceptionally low instabilities. At present a combined instability of the optical frequency standards with cold calcium atoms and an optical comb generator phase locked to a single mercury ion standard of $7 \times 10^{-15}$ in 1~s has been reached \cite{did01}. This value is already close to the theoretical limit for single ion standards, but still two orders of magnitude higher than what can theoretically be achieved with neutral atom standards.

In this letter we show on the example of the calcium standard how the use of ultracold atoms makes neutral atom optical frequency standards competitive in accuracy to the best microwave standards and can improve the stability to unprecedented levels.

For the interrogation of optical transitions of laser-cooled atoms generally a 'separated field excitation' \cite{ram50} in the time domain \cite{bor84, oat99, tre01} with two consecutive pairs of laser pulses from opposite directions is used. 
Such a pulsed excitation scheme represents an atom interferometer with laser pulses acting as beam splitters in analogy to an optical Mach-Zehnder interferometer \cite{bor89}.

The excitation sequence leads to a cosine-shaped signal where the argument of the cosine is given by the phase 
$\Phi = 4 \pi \, T_{\mathrm{sep}} (\nu_{\mathrm{laser}} - \nu_{\mathrm{Ca}})+ (\phi_2-\phi_1)+(\phi_4-\phi_3)$ 
which depends on the time $T_{\mathrm{sep}}$ between the pulses in each pulse pair, the laser detuning 
$\nu_{\mathrm{laser}} - \nu_{\mathrm{Ca}}$ 
and the laser phases $\phi_i$ in the i-th interaction.
These phases appear in this atom-light interferometer because the phase of each beamsplitting laser pulse is transferred to the atomic partial wave.
For perfect alignment these phase differences cancel, however, net phase differences remain when atoms of the cloud move between the pulses to a position where the local phase of the laser is different, e.g. due to curved wavefronts or tilted laser beams \cite{tre01}.

To achieve optimum visibility of the resulting interference fringes, 50 \% splitters are necessary. For resonant excitation this requires $\pi/2$-pulses, defined by the Rabi angle $\Omega_R \times \tau = \pi/2$ with the Rabi frequency $\Omega_R \propto \sqrt{I \gamma}$. 
In particular for narrow linewidth $\gamma$, as needed in a frequency standard, the Rabi frequency at the available laser intensity $I$ is rather low, setting a lower limit on the pulse duration $\tau_p$ which is $\tau_p \approx 1~\mu$s in our experiment.
For non-resonant excitation, the excitation probability is diminished with a half width equal to the Fourier width of the single pulse. Due to the Doppler effect this translates to an acceptance range of atomic velocities, that effectively contribute to the interference signal. For our set up this acceptance range is in the order of 15 cm/s. For broader velocity distributions only a part of the atomic ensemble contributes to the atom interferences, thereby reducing the contrast and the SNR.

In addition, the velocity dependent excitation probability influences the amplitude and the incoherent background of the observed signal. This background may shift the central interference fringe from the undisturbed line center. To correct for this shift and to recover the undisturbed center the signal has to be calculated by an integration over the actual velocity distribution, which becomes less reliable as the velocity spread becomes wider than the acceptance range.

We realize a neutral atom optical frequency standard by stabilizing a laser ($\lambda = 657$~nm) to the intercombination transition from the ground state $^1S_0$ to the metastable state $^3P_1$ of $^{40}$Ca which has a natural linewidth of $\gamma/2 \pi = 320$~Hz \cite{pas80}. Like other alkaline earth elements, $^{40}$Ca has a $^1S_0$ ground  state that is little sensitive to external fields, making it an ideal candidate  for an optical frequency standard, and a nearly closed dipole transition $^1S_0$--$^1P_1$ for effective laser cooling. About 10$^7$ atoms at a temperature of a few millikelvin are prepared by means of a magneto-optical trap (MOT). The corresponding velocity width of this cold ensemble previously presented a major contribution to the uncertainty of the Ca standard of $2 \times 10^{-14}$ \cite{wil02} as a result of the influences mentioned above. 

To overcome this limitation we apply a new quench cooling and trapping scheme on the narrow intercombination transition \cite{bin01a} leading to ultracold atoms with a temperature below $T = 10~\mu$K. 

\begin{figure}
\includegraphics[width=8cm]{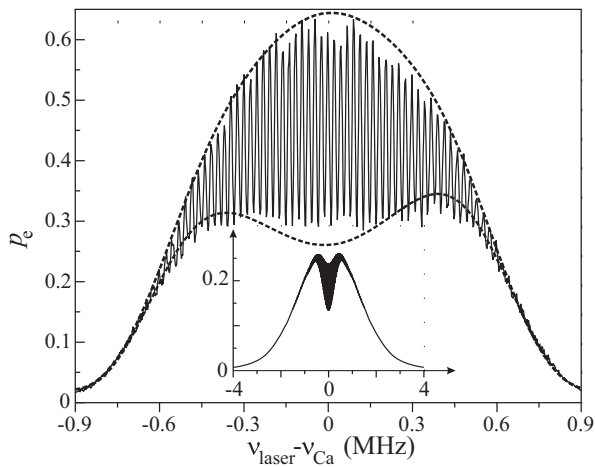}
\caption{Excitation probability of an ensemble of ultracold Ca atoms using a four pulse interferometer. Solid line: fluorescence of the ensemble of Ca-atoms with $v_\mathrm{rms}$ = 7.5 cm/s, $\tau_p$ set to 1.1 $\mu$s. Dashed line: calculated envelope with the above values and $\Omega_{R} \times \tau_p = \pi / 2 $. The inset shows the calculated signal for an atomic sample of $T = 2.7$~mK.}
\label{FullRamsey}
\end{figure}
Now, the frequency width of the whole signal is given by the Fourier width of the exciting pulses (fig.~\ref{FullRamsey}) which is dramatically different from the signal of cold atoms ($T = 2.7$~mK, inset of fig.~\ref{FullRamsey}) where the width is given by the Doppler width.
Therefore, with the strongly reduced width of the velocity distribution nearly all atoms contribute to the interference signal, leading to an observed contrast of 0.36, i.\,e. 70~\% of the optimum contrast that would be achievable with atoms at zero velocity. The interrogation is performed with a ballistically expanding ensemble of ultracold Ca atoms released from the MOT after 220~ms of cooling, where the interference pattern is measured by detecting the fluorescence for 500~$\mu$s emitted in the decay of the $^3P_1$ state after the last pulse of the atom interferometric sequence that takes between 45~$\mu$s and 1.3~ms resulting in a single measurement cycle time $T_\mathrm{cyc} \approx 220$~ms.
The measured interference pattern is in good agreement with calculations based on a spinor approach \cite{bor84}. Here a Gaussian velocity distribution and rectangular pulses were assumed resulting in a contrast $K = 0.42$.
The asymmetry in the line shape originates from the sum of the excitation probabilities of two interferometers realized at the same time by the pulsed excitation scheme. Due to the photon recoil the two interferogramms are centered at $\nu_{\mathrm{Ca}} \pm \mathrm{11.5~kHz}$ while the incoherent 
background is centered at $\nu_{\mathrm{Ca}} + \mathrm{11.5~kHz}$. The small deviations from theory can be attributed to slow variations of atom number during the long scanning time of 40 minutes with 720 points averaged over 15 measurement cycles each.


The SNR of the detection scheme used so far is limited by fluctuations in the atom number and by the shot noise of the detected fluorescence photons, because only one photon per excited atom is emitted and from those only a small fraction can be detected.
The ultimate instability $\sigma_y$ of a frequency standard as a function of the averaging time $\tau$ is given by the quantum projection noise (QPN) limit \cite{ita93} to
\begin{equation}
\sigma_{y}(\tau) = \frac{1}{\pi} \times  \frac{1}{\nu_\mathrm{Ca}\,4T_\mathrm{sep}} \times \sqrt{\frac{1-\overline{p}}{N_0 K^2\overline{p}}} \times \sqrt{\frac{T_\mathrm{cyc}}{\tau}}.
\label{SigNuQPN}
\end{equation}
The second term is the inverse of the quality factor $Q=\nu_\mathrm{Ca}/\delta\nu$ with the effective linewidth $\delta\nu$ set by the pulse separation $T_\mathrm{sep}$. The third term is the QPN-limited SNR of one measurement cycle depending on the contrast $K$, the mean excitation probability $\overline{p}$ and the total number of atoms $N_0$. %
 
\begin{figure}
\includegraphics[width=8cm]{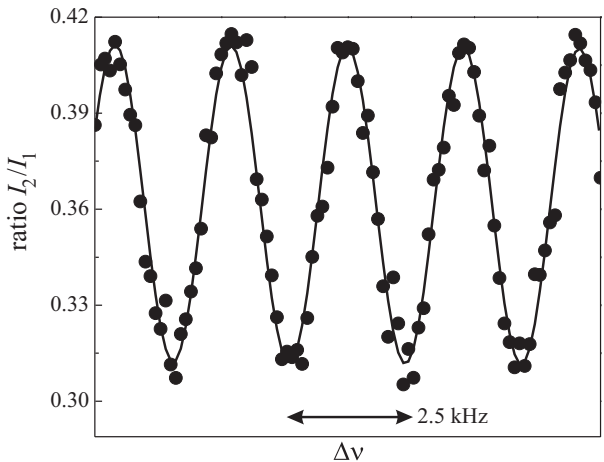}
\caption{Atom interferences detected by the state-selective method. Shown is the intensity ratio of scattered blue light after the decay of $^3P_1$ state atoms ($I_2$) 
to the fluorescence immediately after the atom interferometry ($I_1$) without averaging as function of the laser detuning $\Delta\nu$. Due to detection laser beam direction a strong background fluorescence from atoms of the thermal beam was present.}
\label{BlueRamsey}
\end{figure}

To reach the QPN-limit a detection probability close to unity is necessary. It can be obtained by detecting ground state atoms via their resonance fluorescence when cycled on the cooling transition in the singlet system, which gives a high number of photons per atom. However, such electron shelving detection schemes \cite{nag86} for optical frequency standards with cold atoms have suffered from atom number fluctuations or from heating of the atoms when a normalization was applied \cite{oat99}. 
The novel scheme presented here measures both the ground and the excited state atoms {\it after} the interferometry. Immediately after the last pulse of the atom interferometric sequence the number of atoms in the ground state is measured by cycling them on the $^1S_0$--$^1P_1$ transition by a resonant single laser beam and detection of the fluorescence. By the radiation pressure the atoms are accelerated and should leave the interaction region.
After one to two natural lifetimes of the $^3P_1$ state this measurement is repeated, giving a measure for the number of atoms that have been in the excited state after the interferometry. In addition to waiting for the spontaneous decay also a pulse on the quenching transition was applied. From both values a normalized excitation probability independent of atom number fluctuations can be derived. 
This scheme is particularly suited for ultracold atoms due to the low average velocity and the slow expansion of the atomic cloud by less than 50 $\mu$m, while for a cold ensemble, part of the excited atoms would have left the detection region before decaying to the ground state. Our new simple scheme avoids any heating of the ensemble before the spectroscopy and allows the elimination of atom number fluctuations. Hence, no basic limitation exists any longer that prevents one from reaching the quantum projection noise limit. Fig.~\ref{BlueRamsey} shows interference fringes with a period of 2.3~kHz that were detected with the new state-selective scheme probing an ensemble of about $10^6$ atoms. The SNR is increased by a factor of 6 compared to the SNR reached with the previous detection method. The improvement corresponds to an instability of $5 \times 10^{-14}$ in 1~s and was mainly limited by amplitude noise of the detection laser.

For the principle limit of the stability  we calculated a $\sigma_y(1~s) = 4 \times 10^{-17}$ according to eq.~\ref{SigNuQPN} based on the data from our experimental set up. Here we took into account the improvements of our experimental set up with the cooling laser power increased by an order of magnitude, the utilization of a Zeeman slower and the transfer efficiency to the ultracold ensemble of 12~\% increased to an optimal value of 50~\%. With these improvements we expect $N_0=3 \times 10^7$ atoms in a cycle time $T_\mathrm{cyc}=20$~ms with the temperature at the limit of the quench-cooling $T=6~\mu$K~\cite{bin01a}. We find an optimum stability at a resolution of 390 Hz with a corresponding contrast of $K=0.32$ and mean excitation probability $\overline{p}=0.2$ where we take into account the degradation of coherence in the atom interferometer due to the duration of the interferometry ($\sim\!\! 2 T_\mathrm{sep}$) approaching the lifetime of the excited $^3P_1$ state and the finite pulse width in combination with the Doppler broadening of the ensemble. 

To reach the QPN-limit for our set up, an intensity stability of the detection laser of $10^{-4}$ and a frequency stability of 300~kHz is sufficient. For the spectroscopy laser an instability of $3 \times 10^{-16}$ is necessary for the duration of the atom interferometry $2 T_\mathrm{sep} \approx 1.3$~ms. While laser instabilities better than 1~Hz in 1~s have already been demonstrated~\cite{sal88,you99}, to have the necessary performance of the free running spectroscopy laser at one's disposal is still a challenging task.  

\begin{figure}
\includegraphics[width=8cm]{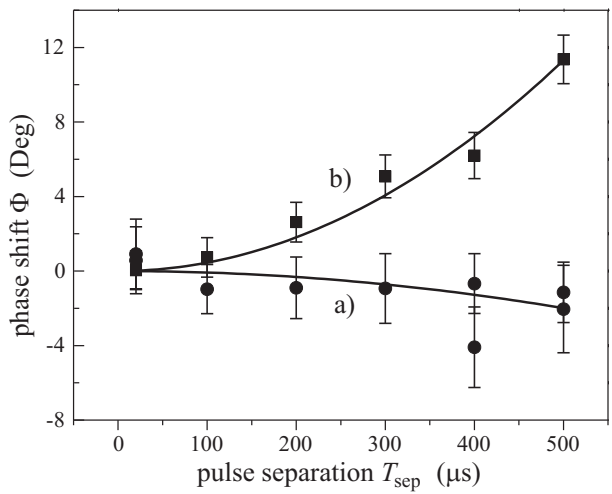}
\caption{Shift of the interferogram of a symmetric three-pulse interferometer
with an intentionally misaligned laser beam.
a) With ultracold atoms ($v_\bot =7.5$~cm/s) the measured phase shift is determined by the deviation of 0.09 degrees from horizontal alignment.
b) With an ensemble at 2.8 mK ($v_\bot = 1.1$~m/s) the gravitational
contribution (the same as in a)) is hidden by the contribution of the wavefront curvature.}
\label{Phaseys}
\end{figure}

In addition to high stability the utilization of ultracold atoms leads to a considerably increased accuracy.
The main contribution to the uncertainty of the Ca frequency standard arises from residual first order Doppler effects due to wavefront imperfections of the exciting laser beams.  
E.g., in a symmetric three-pulse atom interferometer with equal pulse separations $T_{\mathrm{sep}}$ the movement of the atoms through the wavefronts (radius of curvature $R$) of the exciting laser beam results in a phase shift
\begin{equation}
\Phi = A \,T_{\mathrm{sep}}^2
\;\;\mathrm{with}\;\; 
A = \vec {k} \vec{g} + |\vec{k}| v_{\bot}^2 / R , 
\end{equation}
where $\vec{g}$ denotes the gravitational acceleration and $v_{\bot}$ the velocity component perpendicular to the wavevector of the exciting laser pulse $\vec{k}$.
For an intentionally misaligned beam ($R \approx 12$~m) this phase shift in dependence of the time between the pulses is shown in fig.~\ref{Phaseys} for a 
Doppler-cooled ensemble ($T$ = 2.8 mK) and for a quench-cooled ensemble ($T$ = 14 $\mu$K). Because of the reduction of $v_{\bot}^2$ by a factor of more than 200, the second term in the above equation becomes negligible and the remaining phase shift can be fully explained by the slight deviation from horizontal alignment of the laser beam ($\vec{k}\vec{g} \neq 0$) while for the Doppler-cooled ensemble this effect is hidden by the much stronger influence of the wavefront curvature.
By measuring  the quadratic dependencies for both directions that are used in the four pulse interferometer with antiparallel laser beams, the major contribution of the Doppler effect to the frequency shift that changes linearly with $T_\mathrm{sep}$ can be calculated and corrected for.
We investigated how to control and optimize the wavefront curvature with atom interferometric as well as optical means. The combination of optimized wavefronts ($|1/R|<1/300$~m$^{-1}$, beams perpendicular to gravity within 6~$\mu$rad and antiparallel within 22~$\mu$rad) together with the ultracold atoms ($T = 6~\mu$K, average velocity $< 1$~mm/s) will reduce the contribution of the residual Doppler effect to the uncertainty from 4.6~Hz to only 150~mHz \cite{wil02,wil02z}.

At this level other contributions like collisions between the atoms will play the dominant role. 
From our measurements at $T\approx 3$ mK we determine the coefficient of the density-dependent relative frequency shift $\alpha = (3.0 \pm 4.4 )\times 10^{-30} \,\mathrm{m^{3}}$ \cite{wil02} which is more than a factor of 200 smaller than in the Cs microwave clock.
With the accuracy and stability now at hand using ultracold atoms this dependence can be measured more accurately and then be used to correct the clock frequency accordingly. Assuming the same coefficient $\alpha$ for the ultracold atoms and a similar density of $1.3 \times 10^9$~cm$^{-3}$ known with a relative uncertainty of 10~\% we estimate the uncertainty for the correction to be 260~mHz. Altogether, the expected total relative uncertainty will be $8 \times 10^{-16}$ with other contributions due to magnetic fields (80~mHz), blackbody radiation (50~mHz) and electric fields (20~mHz). From calculated interferograms (Fig.~\ref{FullRamsey}) we estimate the uncertainty due to the line asymmetry for the application of a 3f-stabilization scheme at a resolution of 390~Hz to below 50~mHz with no correction applied \cite{wil02,wil02z}.

In conclusion, we presented measurements that show how the calcium optical frequency standard can reach a relative uncertainty of below $10^{-15}$ 
and a quantum projection noise limited instability of $ 4 \times 10^{-17}$ in only 1~s, which will make it competitive in accuracy and superior in stability to existing microwave standards and single ion optical clocks, respectively. Besides the use as an optical clock the sensitivity, the high signal-to-noise ratio and the reliable theoretical modeling of the atom interferences can be used for various sensor applications. 

The work was supported by the Sonderforschungs\-be\-reich SFB 407 of the Deutsche Forschungs\-gemeinschaft and by the PhD programme of the 
 Physikalisch-Technische Bundesanstalt. 


\end{document}